# Two Procedures for Robust Monitoring of Probability Distributions of Economic Data Streams induced by Depth Functions

Daniel Kosiorowski, Department of Statistics, Cracow University of Economics
Cracow, Poland, e-mail: daniel.kosiorowski@uek.krakow.pl

**Abstract:** Data streams (streaming data) consist of transiently observed, evolving in time, multidimensional data sequences that challenge our computational and/or inferential capabilities. In this paper we propose user friendly approaches for robust monitoring of selected properties of unconditional and conditional distributions of the stream based on depth functions. Our proposals are robust to a small fraction of outliers and/or inliers, but at the same time are sensitive to a regime change in the stream. Their implementations are available in our free R package DepthProc.

**Keywords:** Data Stream; Robust Procedure; Statistical Depth Function

## I. INTRODUCTION

The amounts of data nowadays at economists disposal, force them to use different decision algorithms to those used merely several years ago. On-line credit scoring, intrusion detection to the computer systems or algorithmic future contracts trading are examples of phenomena which have initiated the evolution of data analysis techniques and methods of statistical inference (see [10] for provoking overview of challenges which statisticians and econometricians face due to modern on-line science and trading).

The main motivation of this paper relates to a new phenomenon which has appeared in the economics literature in recent years called **data stream analysis (DSA)** (or streaming data processing). This terminology originates from theoretical informatics (see [1], [20]). Generally speaking, *in the case of DSA we have to cope with huge amounts of constantly updated data that enter, at non-equally spaced time points, a processing system, and we have restricted memory and computational resources for processing the data*. We are looking for sufficient, as well as computationally and memory tractable, statistics for the issues under our consideration. The algorithms applied in DSA have to fulfil strict criteria in the context of **1**. the speed of data transmission to a program, **2**. computational complexity of the algorithm, **3**. amount of memory necessary to apply the algorithm. An algorithm should be highly elastic in adaptation to changes in the data generating mechanism (see [2]).

Although DSA originates from informatics, modern time series econometrics deals with similar research problems, for example in the context of analysing multivariate financial time series or studying sales data. **Economic data streams** additionally consist of a small or a moderate fraction of outliers or inliers of various kinds. From one point of view, we can say, *that economic data stream analysis (EDSA) involves using locally sufficient statistical procedures which are robust to outliers and inliers, but at the same time sensitive to the data generating mechanism, and which are computationally tractable to a degree allowing their use on-line.* From the other perspective, *the EDSA mainly deals with detecting outliers, which contain useful information related to e.g., a credit card fraud detection, law enforcement or certain intrusion detection.*

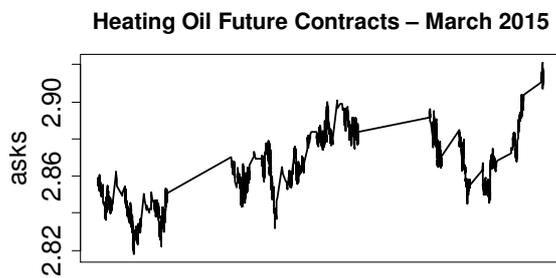

**Fig. 1** Heating oil future contracts – asks.

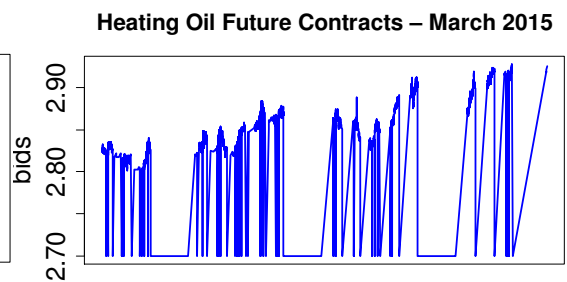

**Fig. 2** Heating oil future contracts – bids.

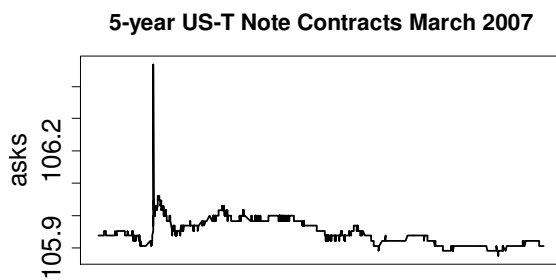

**Fig. 3** 5-year US-T note contracts – asks.

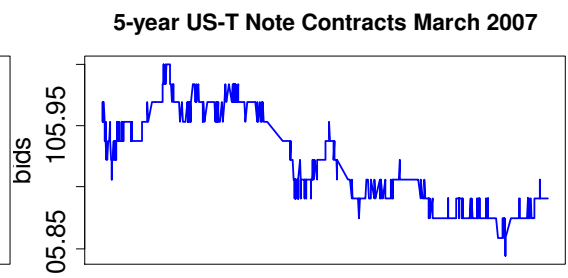

**Fig. 4** 5-year US-T note contracts – bids.

Fig. 1 – 4 present order book data on future contracts for heating oil and 5-years US – T note contracts. These series may be treated as examples of economic data streams, i.e., observations appear at non-equally spaced time points, the data contain isolated outliers and paths of outliers, the data generating mechanism evolves rapidly in time. Note that, in situations like these, we cannot use popular moving average or ARIMA modelling setup for prediction purposes (data are not equally spaced, the underlying process is not stationary).

In recent years several approaches to EDSA have been proposed. These proposals involve *parallel computing* and, *hierarchic algorithms* for arriving data. It is worth noticing, that in the context of EDSA, we can observe a renaissance of well–known simple statistics, which can be calculated using *recursive* and/or *parallel* calculation. We can list here: *mean vectors*, *covariance matrices*, *dynamic least squares regression*, the *Kalman filter* (see [14], [15]).

In this paper we consider selected issues related to the robust monitoring of conditional and unconditional distributions of a data stream. Note, that in time series econometrics, the distribution conditioned on the observed past is called the *predictive distribution*.

The rest of the paper is organized as follows. In Section 2, we briefly outline recent developments in the area of estimating the conditional distribution. In Section 3, we present a general framework for our considerations. In Section 4, we give a brief review covering elements of the concept of data depth. In Section 5, we propose two depth based strategies for online monitoring of conditional and unconditional data stream distributions. In Section 6, we discuss properties of the proposed strategies. Finally, we present concluding remarks in Section 7. The paper ends with references.

## II. SELECTED ASPECTS OF ESTIMATING A CONDITIONAL DISTRIBUTION

One aspect of estimating and monitoring a conditional distribution (CD) relates to the following basic question: *does observation of the past increase our abilities to react to future events*. Monitoring a CD is related to detecting investment opportunities, construcion effective investment strategy, obtain insight into the relation between the future and the present of a certain economic phenomenon. The CD may be determined by a cumulative distribution function, density function or imprecisely (in general) by a certain set of descriptive measures.

Let us consider a random vector $(X, Y)$ with *cumulative distribution function* (cdf) $F(x, y)$. We wish to estimate the **conditional distribution function** of $Y$ given $X = x$,

$$F_x(y) = P\{Y \leq y \mid X = x\}, \quad y \in \mathbb{R}. \tag{1}$$

In a typical setting, we have equally spaced in time i.i.d observations $(X_1, Y_1), ..., (X_n, Y_n)$ from a random vector $(X, Y)$. Our goal is to estimate $F_x(y)$ based on these observations. Recalling a representation of the conditional cdf at a point as a regression of an indicator variable (see [25], [26])

$$F_x(y) = P\{Y \leq y \mid X = x\} = \mathbb{E}\{\mathbf{1}\{Y \leq y\} \mid X = x\},$$

where $\mathbf{1}\{\cdot\}$ is an indicator function, we can use a natural nonparametric estimator of (1)

$$\hat{F}_x(y) = \sum_{i=1}^{n} w_{ni}(x, h_n) \mathbf{1}\{Y_i \leq y\}, \qquad (2)$$

where $\{w_{ni}(x, h_n)\}$ denotes a certain sequence of weights (e.g. Nadaraya-Watson weights or local linear weights) with $h_n > 0$ being a bandwidth sequence, $h_n \to 0$ as $n \to \infty$.

This approach was studied in [7] and [8], among others, and recently was presented in [6].

Following [6], there are four general categories of approaches to estimating the conditional cdf:

1. *Fully nonparametric approach,* where we have no assumption about the effect of the covariate X on the variable Y;

2. *Parametric approach,* where the conditional distribution function can be expressed as $F_x(y, \boldsymbol{\theta})$, for a certain vector of parameters $\boldsymbol{\theta}$;

3. *Semiparametric location-scale* model:
$$Y = m(X, \theta) + \sigma(X, \theta)\varepsilon, \qquad (3)$$
where $m(\cdot)$ and $\sigma(\cdot)$ are known functions and the distribution of $\varepsilon$ is unknown, as well as;

4. *Nonparametric location-scale* model
$$Y = m(X, \theta) + \sigma(X, \theta)\varepsilon, \qquad (4)$$
where $m(\cdot)$ and $\sigma(\cdot)$ are unknown smooth functions, and the distribution of $\varepsilon$ is unknown.

The estimators applied to DSA may belong to any of the above categories. However, they should be computationally tractable, robust to a small fraction of outliers and inliers and possess what might be called *forgetting mechanism* enabling adaptation to a change in the data generating mechanism regime (see [2]). The simplest forgetting mechanism may be obtained by introducing estimation basing on a ***moving window*** of a fixed, random or data driven length.

Let $\{X_1, X_2, ...\} \subset \mathbb{R}^d$ be an economic data stream, $d \geq 1$. A **window** $\mathbf{W}_{i,n}$ denotes the sequence of points of the stream ending at $X_i$ of size $n$, i.e., $\mathbf{W}_{i,n} = (X_{i-n+1}, ..., X_i)$, where $i \in I_1 = \{1, 2, ...\}$ or $i \in I_K = \{K, 2K, 3K, ...\}$, $K \in \mathbb{N}$. We make a decision at moment $i+1$ based on the information contained in a fixed number of windows $\mathbf{W}_{i_1, n_1} \in \mathcal{W}_{i_1, n_1}, ..., \mathbf{W}_{i_K, n_K} \in \mathcal{W}_{i_K, n_K}$, $i_1 \in I_1, ..., i_K \in I_K$, $n_1 < ... < n_K$ ($\mathcal{W}_{i_1, n_1}$ denotes all the collections of linear combinations of elements in $\mathbf{W}_{i_1, n_1}$ — all the information contained in the window of length $n_1$ available at time $i_1$).

For further considerations it is useful to introduce a fixed **number of $r$ reference windows** $\mathbf{W}_1^r, ..., \mathbf{W}_M^r$ related to our prior knowledge about the $M-$ regimes of the stream or related to various decision criteria. We assume that the reference windows $\mathbf{W}_j^r$ are constant over time or are updated with significantly smaller frequency than the "working" moving window $\mathbf{W}_{i,n}$, which is the main tool for the stream analysis (say weeks and minutes, respectively).

Monitoring the stream, we should note that for a stream consisting of several regimes – a point which is outlying with respect to (w.r.t) one regime may not be outlying w.r.t. another regime. The procedure applied to the DSA should be robust, but not very robust – it should be sensitive to regime changes, but unaffected by outlying points at the same time.

In the "*classical*" setting of CD estimation, several authors proposed various improvements of the estimator (2) appealing to the general idea of making some pre-adjustment, inspired by some specific model structure, but without assuming that this model structure holds.

Suppose that the relation between the present and the past may be expressed by means of a simple linear regression model $Y_i = \theta_0 + \theta_1 X_i + \varepsilon_i$, $i = 1, ..., n$, where $\varepsilon_1, ..., \varepsilon_n$ are i.i.d. with the same distribution as $\varepsilon$ (denoted by $F_\varepsilon$ and unknown).

The CD of $Y$ given $X = x$ is thus

$$F_x(y) = P\{Y \leq y \mid X = x\} = P\{\theta_0 + \theta_1 X + \varepsilon \leq y \mid X = x\} = F_\varepsilon(y - \theta_0 - \theta_1 x) . \tag{5}$$

If the simple regression model holds, we can estimate $F_x(y)$ by

$$\hat{F}_\varepsilon(y - \hat{\theta}_0 - \hat{\theta}_1 x) = \sum_{i=1}^{n} \frac{1}{n} \mathbf{1}\{\hat{\varepsilon}_i \leq y - \hat{\theta}_0 - \hat{\theta}_1 x\} , \tag{6}$$

where $\hat{F}_\varepsilon$ is the empirical distribution function of the residuals, $\hat{\varepsilon}_i = Y_i - \hat{\theta}_0 - \hat{\theta}_1 X_i$

Please note, that for robust DSA we can use robust estimators for simple regression, such as the M- estimator, least trimmed squares estimator or deepest regression estimator (see [19]).

**In another family of approaches,** data are pre-adjusted by some location and scale model

$$m(x) = E[Y \mid X = x] \text{ and } \sigma^2(x) = E[Y^2 \mid X = x] - m^2(x) , \tag{7}$$

We transform the observations $Y_i$ into $Y_i^a$, where

$$Y_i^a = \frac{Y_i - m(X_i)}{\sigma(X_i)} ,$$

or when the location and scale parameters need to be estimated

$$\hat{Y}_i^a = \frac{Y_i - \hat{m}_n(X_i)}{\hat{\sigma}_n(X_i)}.$$

This approach leads to the nonparametric pre-adjusted estimator

$$\hat{F}_x^a(y) = \sum_{i=1}^n w_{ni}(x, h_n) \mathbf{1}\left\{ \frac{Y_i - \hat{m}_n(X_i)}{\hat{\sigma}_n(X_i)} \leq \frac{y - \hat{m}_n(x)}{\hat{\sigma}_n(x)} \right\}, \qquad (8)$$

Typical examples of estimators for $m(\cdot)$ and $\sigma^2(\cdot)$ are local linear regression estimates. Using pre-adjustment of the data, we hope to reduce the bias of the estimator (while not increasing its variance).

The CD may be determined as well by its *conditional density function*. In our opinion, the best proposal in this context is the conditional density estimator based on the local linear approximation proposed and studied in Hyndman & Yao [11] and implemented in the hdrcde R package [12].

Figures 5 – 8 relate to 5-min quotations of stocks belonging to the Dow Jones Industrial Index in the period from 2008-03 to 2013-03. This period was divided into 5 consecutive sub-periods of equal length. Figures 5–8 present estimated predictive distributions of the present value conditioned on the past values for one stock *Catepillar Inc*. belonging to the Dow Jones Industrial Index. The distributions were estimated using an estimator implemented within the hdrcde R package. It is easy to notice general changes in the shapes of these distributions, but after a thoughtful look we can conclude that the distribution evolves from a bimodal distribution in a period of stagnation to multimodal in the period of Greek crisis and back to bimodal in the next periods. We hope that the proposed strategies for robust analysis of the PD can provide a clearer picture of this evolution and also be conducted online, which would be useful in predicting future crashes.

The methods outlined above are computationally very intensive and due to the so called *curse of dimensionality* (see [3], [22]), they perform relatively poorly in high dimensional problems. Issues related to their robustness have not yet been well developed. Therefore, we further propose a simple but powerful method of robustly decreasing the complexity of the estimation for monitoring one-dimensional streams by appealing to the well-known idea of binning (see [25], [26]) and we propose using multivariate Wilcoxon type statistics in the case of monitoring multivariate data streams.

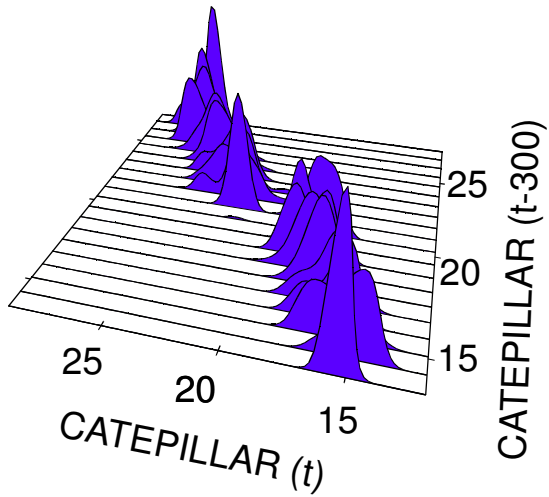 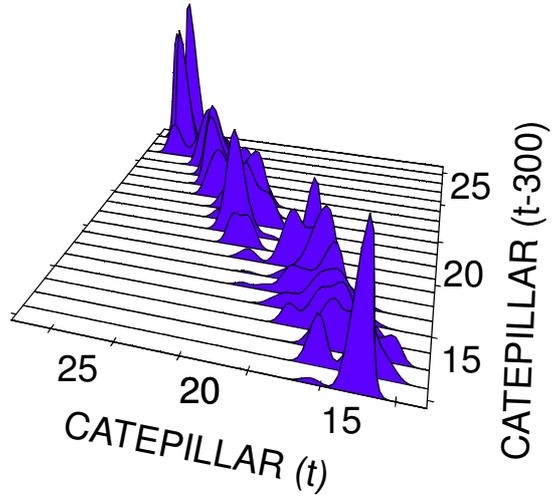

**Fig. 5** The estimated PD 2008-03 to 2009-03.
*Source*: Calculations using the hdrcde R package.

**Fig. 6** The estimated PD 2009-03 to 2010-03.
*Source*: Calculations using the hdrcde R package.

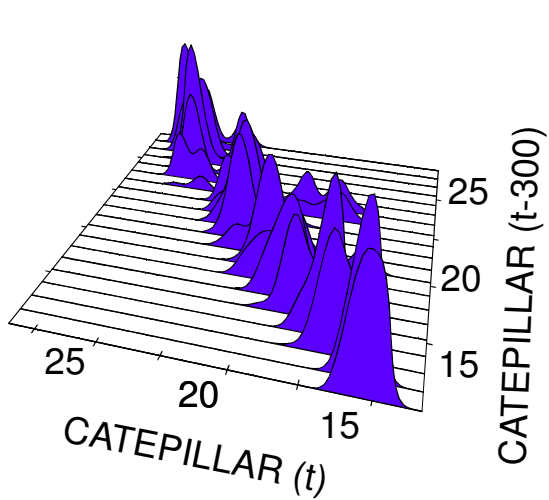 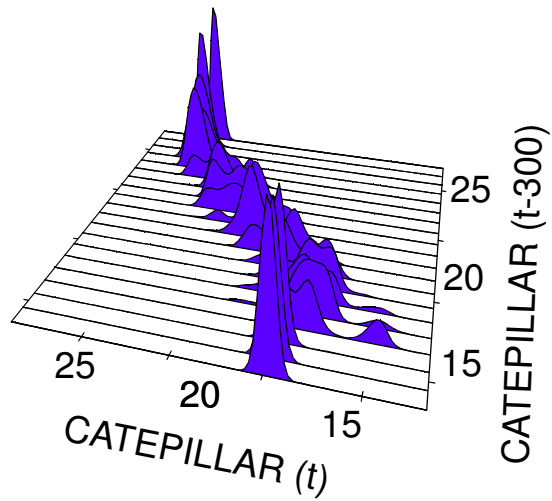

**Fig. 7** The estimated PD 2010-03 to 2011-03.
*Source*: Calculations using the hdrcde R package.

**Fig. 8** The estimated PD 2011-03 to 2012-03.
*Source*: Calculations using the hdrcde R package.

### III. MODEL ENOMIC DATA STREAM AND RESEARCH ISSUES

One of the main features of economic data streams, relates to changes in their regimes. The stochastic characteristics of a particular regime of the stream may be treated as a vocabulary, using which a market or social network responds to a certain event, unexpected news or government intervention. For a description of the uncertainty related to data stream analysis, it seems natural to use one of the multi-regime time series models known in the econometric time series literature (see [24]). For our purposes, we propose to make use of two general schemes, respectively representing random and deterministic switching between regimes – the CHARME (*conditional heteroskedastic autoregressive mixture of experts*) for random switching between regimes (for details see [23]) and SETAR (*self-exciting threshold autoregressive model*) ( see [24]).

According to the **CHARME** model, a hidden Markov chain $\{Q_i\}$ with a finite set of states $\{1, 2, ..., M\}$ drives the dynamics of the stream $\{X_i\}$:

$$X_i = \sum_{j=1}^{M} S_{ij}(m_j(X_{i-1}, ..., X_{i-p}) + \sigma_j(X_{i-1}, ..., X_{i-p})\epsilon_i), \qquad (9)$$

with $S_{ij} = 1$ for $Q_i = j$ and $S_{ij} = 0$ otherwise, $m_j$, $\sigma_j$, $j = 1, ..., M$, are unknown functions and $\epsilon_i$, $i = 1, 2, ..., n$ are i.i.d. random variables with mean zero. To be able to conduct statistical inference, we assume that $Q_i$ changes its value only rarely, i.e., the observed process follows the same regime for a relative long time before any change in the regime occurs.

The properties of and conditions for the geometric ergodicity of the model (9) are given in [23]. It is worth noticing that in the case of a mixture of $M > 1$ regimes, the stationary conditions for (9) do not have to hold for all the states, but only for those which are frequently visited. This is especially interesting in the context of modelling economic streams – we very often observe economic phenomena, where non-stationary periods of panic (involving a major revision of predictions of the future), appear, but do not dominate any general stationary tendency. The CHARME model represents a random switching scheme.

Our second proposal for the modelling of streaming data, concerns a relatively popular model in the econometrics literature model with deterministic switching called SETAR, which assumes deterministic switching time. For a one-dimensional time series $\{X_t\}$, a SETAR model of order $p$ is defined by:

$$X_t = \sum_j \mathbf{1}_{A(j)}(z_t)(b_{0j} + b_{1j}X_{t-1} + ... + b_{pj}X_{t-p}), \qquad (10)$$

where $A_1, ..., A_M$ denotes some finite partition of the real line, $Z_t$ is a variable, depending on which level a change in the regime occurs, usually $Z_t$ is one of the lagged variables $\{X_{t-1}, ..., X_{t-p}\}$, $\mathbf{1}_A(x)$ denotes the indicator function taking value 1 for $x \in A$ and 0 in other cases.

The SETAR model describes an asymmetry in how a process increases and decreases, as observed in practice. It uses piecewise linear models to obtain a better approximation of the equation for the conditional mean. However, in contrast to the traditional piecewise linear model that allows model changes to occur in the „time" space, the SETAR model uses a threshold in "space" to improve the linear approximation. Under the SETAR model, a transition between the regimes is determined by a particular lagged variable. Consequently, the SETAR model uses a deterministic scheme to govern transitions between regimes. Under the CHARME model, a stochastic scheme related to a hidden Markov Chain rules the regime changes. In practice, the stochastic nature of states implies that one is never certain about which state $x_t$ belongs to under the CHARME model. This difference has important practical implications in forecasting. For instance, classical econometric forecasts using the CHARME model are always a linear combination of those of forecasts produced by sub-models of individual states. But those obtained using the SETAR model only come from a single regime, provided that $x_{t-p}$ is observed. Forecasts under a SETAR model are also based on a linear combination of those produced by models of individual regimes when the forecast horizon exceeds the delay $p$. It is much harder to estimate using the CHARME model rather than other models because the states are not directly observable.

To take into account that in the case of DSA observations are not equally spaced and the intensity of observations may vary, models (9) or (10) may be additionally associated with a certain Poisson process. We develop this issue elsewhere. *Economic data streams* usually consist of a moderate fraction of *outliers* or *inliers*, i.e., instead of observing $X_i$ we observe $Y_i = X_i + b_i \theta_i$, where $b_i \theta_i$ represents an additive outlier term (a point which in some sense departs from the majority of the data) or an inlier (a point which artificially increases the degree of multimodality of the data), $b_i$ is an unobservable binary random variable $P(b_i = 0) = 1 - \varepsilon$, $\varepsilon -$ denotes the fraction of outliers or inliers, $\theta_i$ denotes the random magnitude of an outlier or inlier (see [22]).

In the case of a monitoring the **conditional or unconditional distribution of a stream**, we distinguish a finite set $R = \{h_0, ..., h_M\}$ of densities belonging to a certain family $\mathcal{F}$. For a fixed moment $i$, we can treat the elements of $R$ as $M+1$ hypotheses for a test, i.e., any $\mathcal{A}-$measurable function $\psi^i : \mathcal{X} \to \{0, ..., M\}$. Our decision at moment $i$ depends on the value of a certain **minimum distance test statistic**:

$$\psi^i = \arg\min_{0 \leq k < M} d(\hat{g}_n^i, h_k), \tag{11}$$

where "*d*" denotes a distance e.g., a *Kolmogorov* or *Hellinger* distance, $\hat{g}_n^i$ denotes an estimate unconditional or conditional density of interest.

If we have the **reference samples** $\mathbf{W}_1^r,...,\mathbf{W}_M^r$ instead of the **reference densities**, we can estimate the densities using, for example, a kernel or local polynomial method.

Under a given hypothesis $h^* \in R$, for a fixed moment $i$, we are looking for an optimal procedure, i.e., a procedure satisfying

$$\psi_*^i = \inf_{\psi} \max_{0 \leq j \leq M} P_{h_j}\left(\psi \neq j\right), \qquad (12)$$

under the condition $P_{h^*}(\psi(W_{i,n})) \leq \alpha$, $0 < \alpha < 1$, and where *inf* denotes the infimum over all tests.

The **monitoring of a data stream** comprises of a **sequence of tests** conducted at consecutive moments. Therefore, we are looking for a procedure minimizing criterion (12) over a certain horizon $i \in \{1,...,T\}$, i.e., we are searching for

$$\psi_{**} = \inf_{\psi_*} \sum_{i=1}^{T} \psi_*^i. \qquad (13)$$

Appropriate choices of the distance and density estimator appearing in (11) are a crucial issue related to the quality of the analysis of a stream distribution. Using the *Hellinger* distance, we take into account the "integrated behaviour" of the distribution over the whole of its support, whereas using the Kolmogorov distance, we underline the worst behaviour at a point of the support. The general theoretical framework for studying (11), (12) and (13) in the i.i.d. case may be found in [25], [26].

### IV. DSA TOOLS INDUCED BY THE DATA DEPTH CONCEPT

**The data depth concept** (DDC) was originally introduced as a way to generalize the concepts of the median and quantiles to the multivariate framework. A depth function $D(\cdot, F)$ ascribes to a given $\mathbf{x} \in \mathbb{R}^d$ a measure $D(\mathbf{x}, F) \in [0,1]$ of its centrality w.r.t. a probability measure $F \in \mathcal{P}$ over $\mathbb{R}^d$ or w.r.t. an empirical measure $F_n \in \mathcal{P}$ calculated from a sample $\mathbf{X}^n = \{\mathbf{x}_1,...,\mathbf{x}_n\}$. The larger the depth of $\mathbf{x}$, the more central $\mathbf{x}$ is w.r.t. to $F$ or $F_n$. The best known examples of depth functions in the literature are Tukey and Liu depths (for further details see [16]). Although the DDC offers a variety of user-friendly and powerful tools, it is not well known to a wider audience. These tools are of special value in the context of DSA and in general for multivariate economics.

Thinking in terms of an influential majority of multivariate objects concentrated around the center relates robust statistics for example with welfare economics.

In the context of the EDSA we recommend using the **weighted $L^p$ depth**. The weighted $L^p$ depth $D(\mathbf{x}; F)$ of $\mathbf{x} \in \mathbb{R}^d$, $d \geq 1$ generated by a $d$ dimensional random vector $\mathbf{X}$ with distribution function $F$, is defined by

$$D(\mathbf{x}; F) = \frac{1}{1 + \mathbb{E}w(\|\mathbf{x} - \mathbf{X}\|_p)}, \qquad (14)$$

where $\mathbb{E}$ denotes expected value, $w$ is a suitable weight function on $[0, \infty)$, and $\|\cdot\|_p$ denotes the $L^p$ norm. We assume that $w$ is non-decreasing and continuous on $[0, \infty)$ with $w(\infty-) = \infty$, and for $a, b \in \mathbb{R}^d$ satisies $w(\|a+b\|) \leq w(\|a\|) + w(\|b\|)$. Furthermore, in the role of the weight function we use $w(x) = a + bx$, $a, b > 0$. The empirical version of the weighted $L^p$ depth function is obtained by replacing the distribution function $F$ of $\mathbf{X}$ in $\mathbb{E}w(\|\mathbf{x} - \mathbf{X}\|_p) = \int w(\|x - t\|_p) dF(t)$ by its empirical counterpart calculated from the sample $\mathbf{X}^n = \{\mathbf{x}_1, ..., \mathbf{x}_n\}$

$$D(\mathbf{z}, \mathbf{X}^n) = \left[1 + \frac{1}{n} \sum_{i=1}^{n} w\left(\|\mathbf{z} - \mathbf{x}_i\|_p\right)\right]^{-1}. \qquad (15)$$

A point for which the depth takes its maximum is called the $L^p$ **median** (multivariate location estimator), the set of points for which the depth takes a value not smaller than $\alpha \in [0,1]$ is the multivariate analogue of the quantile and is called the $\alpha-$ central region, $D_\alpha(F) = \{\mathbf{x} \in \mathbb{R}^d : D(\mathbf{x}, F) \geq \alpha\}$.

Theoretical properties of this depth were obtained by Zuo in [28]. The weighted $L^p$ depth function in at a point, has a low breakdown point (BP) and unbounded influence function (IF) but on the other hand, the medians based on the weighted $L^p$ depth (multivariate location estimator) are globally robust with the highest BP for any reasonable estimator. The weighted $L^p$ medians are also locally robust with bounded influence functions for suitable weight functions. A *low BP and unbounded IF at a point and a high BP of an estimator of centrality seems to be especially desirable for DSA*. For example, a projection depth with high BP and bounded IF performs worse than the $L^p$ depth in the DSA. Unlike other depth functions and multivariate medians, the weighted $L^p$ depth and medians are easy to calculate in high dimensions. The price for this advantage is the lack of affine invariance of both the weighted $L^p$ depth and medians. The complexity of calculating the weighted $L^p$ depth is $O(d^2n+n^2d)$ and parallel computing procedures may be used (see [28]).

For any $\beta \in (0,1]$ we can define the smallest depth region bigger or equal to $\beta$

$$R^\beta(F) = \bigcap_{\alpha \in A(\beta)} D_\alpha(F), \qquad (16)$$

where $A(\beta) = \{\alpha \geq 0 : P[D_\alpha(F)] \geq \beta\}$.

Fig. 9 and Fig. 10 present a sample $L^2$ depth contour plot and $L^2$ sample depth perspective plot, respectively, obtained using our DepthProc R package.

For two probability distributions $F$ and $G$, both in $\mathbb{R}^d$, we can define a **depth vs. depth plot,** which is a very useful *generalization of the one dimensional quantile-quantile plot*:

$$DD(F,G) = \{(D(\mathbf{z},F), D(\mathbf{z},G)), \mathbf{z} \in \mathbb{R}^d\}. \qquad (17)$$

Its sample counterpart, calculated for two samples $\mathbf{X}^n = \{\mathbf{X}_1,...,\mathbf{X}_n\}$ from $F$ and $\mathbf{Y}^m = \{\mathbf{Y}_1,...,\mathbf{Y}_m\}$ from $G$, is defined as

$$DD(F_n, G_m) = \{(D(\mathbf{z},F_n), D(\mathbf{z},G_m)), \mathbf{z} \in \{\mathbf{X}^n \cup \mathbf{Y}^m\}\}. \qquad (18)$$

A detailed presentation of the DD-plot can be found in [18] or [16]. Fig. 11 presents a DD-plot with a heart-shaped pattern characteristic in the case of differences in location between two samples, whereas Fig. 12 presents a moon-shaped pattern typical in the case of scale differences between samples.

Applications of DD-plots and theoretical properties of statistical procedures using this plot can be found in [18] and [29].

Having two samples $\mathbf{X}^n$ and $\mathbf{Y}^m$ and using any depth function, we can compute depth values for the combined sample $\mathbf{Z}^{n+m} = \mathbf{X}^n \cup \mathbf{Y}^m$, assuming the empirical distribution is calculated based on all observations, or only on observations belonging to one of the samples $\mathbf{X}^n$ or $\mathbf{Y}^m$. For example, if we observe $X_i$'s that depths are more likely to cluster tightly around the center of the combined sample, while $Y_i$'s depths are more likely to occupy outlying positions, then we conclude that $\mathbf{Y}^m$ was drawn from a distribution with a larger scale.

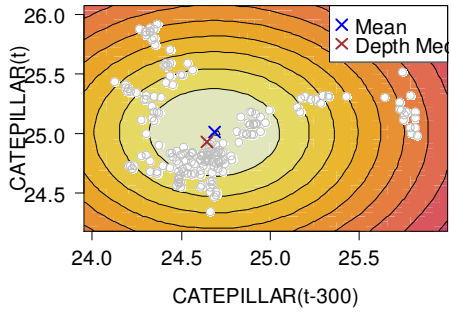
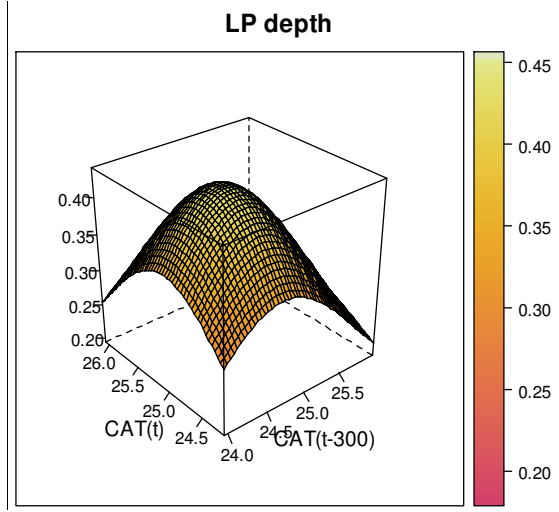

**Fig. 9** Sample $L^2$ contour plot.
*Source: DepthProc R package.*

**Fig. 10** Sample $L^2$ perspective plot.
*Source: DepthProc R package.*

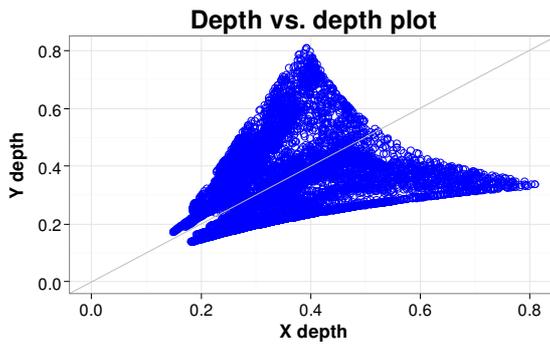
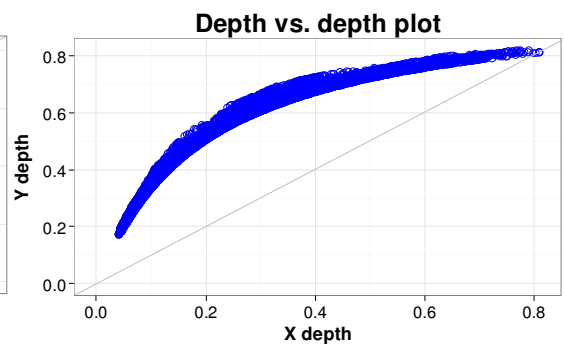

**Fig. 11** Sample DD-plot – location differences.
*Source: DepthProc R package.*

**Fig. 12** Sample DD-plot – scale differences.
*Source: DepthProc R package.*

Properties of DD– plot based statistics in the i.i.d setting were studied in [18]. The authors proposed several DD-plot based statistics and presented "bootstrap arguments" for their consistency and high effectiveness in comparison to Hotelling $T^2$ and multivariate analogues of Ansari-Bradley and Tukey-Siegel statistics.

The depth based **multivariate Wilcoxon rank sum test** is especially useful for the detection of multivariate scale changes and was studied among other in [18].

For the samples $\mathbf{X}^m = \{\mathbf{X}_1,...,\mathbf{X}_m\}$, $\mathbf{Y}^n = \{\mathbf{Y}_1,...,\mathbf{Y}_n\}$, and a combined sample $\mathbf{Z} = \mathbf{X}^n \cup \mathbf{Y}^m$ the **Wilcoxon statistic** is defined as

$$S = \sum_{i=1}^{m} R_i, \tag{19}$$

where $R_i$ denotes the rank of the i-th observation of $\mathbf{X}^m$, $i = 1,...,m$ in the combined sample

$$R(\mathbf{x}_l) = \#\{\mathbf{z}_j \in \mathbf{Z} : D(\mathbf{z}_j, \mathbf{Z}) \leq D(\mathbf{x}_l, \mathbf{Z})\}, l = 1,...,m.$$

The distribution of $S$ is symmetric about $E(S) = 1/2 m(m+n+1)$, its variance is $D^2(S) = 1/12\, mn(m+n+1)$. For the asymptotic distributions of depth based multivariate Wilcoxon rank-sum test statistics under the null and general alternative hypotheses and theoretical properties of such statistics see [29]. Note that using a DD-plot object (implemented in the DepthProc R package), it is easy to calculate other multivariate generalizations of rang test statistics involving, e.g. Haga or Kamat statistics (more sensitive to change in regime) and apply them to the robust monitoring of certain especially interesting features of multivariate time series.

## V. PROPOSALS

Let $(Y, \mathbf{X})$ with $y \in \mathbb{R}$, $\mathbf{x} \in \mathbb{R}^d$ be a random vector with joint density $f(y, \mathbf{x})$ and marginal density of $\mathbf{X}$ $f_X(\mathbf{x})$, then the **conditional density** $g(Y | \mathbf{X} = \mathbf{x}) = f(y, \mathbf{x}) / f_X(\mathbf{x})$, can be estimated by inserting a kernel density, local polynomial or k-nearest neighbors density estimator in both the nominator and denominator of $g(y|\mathbf{x})$. In the context of DSA, $\mathbf{X}$ denotes a vector of lagged values of a phenomenon $Y$. In this case, $g(\cdot | \mathbf{x})$ determines the so called ***predictive distribution*** of $Y$, given $\mathbf{X} = \mathbf{x}$ represents the past.

Let us recall, that **binning** is a popular method enabling faster computation by reducing the continuous sample space to a discrete grid (see [26]). It is useful, for example in the case of estimating a predictive distribution by means of kernel methods. To bin a window of $n$ points $W_{i,n} = \{X_{i-n+1},...,X_i\}$ into a grid $X'_1,...,X'_m$, we simply assign each sample point $X_i$ to the nearest grid point $X'_j$. When binning is completed, each grid point $X'_j$ has an associated number $c_i$, which is the frequency of the points that have been assigned to $X'_j$. This procedure replaces the data $W_{i,n} = \{X_{i-n+1},...,X_i\}$ with the smaller set $W'_{j,m} = \{X'_{j-m+1},...,X'_j\}$. Although simple binning can speed up computation, it has been criticized for a lack of a precise control over the accuracy of the approximation. Robust binning however retains the properties of the majority of the data and decreases the computational complexity of the DSA at the same time.

For a 1D window $W_{i,n}$, let $Z_{i,n-k}$ denote a 2D window created from $W_{i,n}$ consisting of $n-k$ pairs of observations and $k$ lagged observations $Z_{i,n-k} = \{(X_{i-n-k}, X_{i-n+1})\}$, $1 \leq i \leq n-k$.

Further, it is sufficient to consider the simplest case $k = 1$. Assume that we analyze a data stream $\{X_t\}$ using a *moving window* of fixed length $n$, i.e., $W_{i,n}$ and the derivative window $Z_{i,n-1}$. In the first step we calculate the weighted sample $L^p$ depth for $W_{i,n}$. Next we choose an equally spaced grid of points $l_1, ..., l_m$ in such way that $[l_1, l_m] \times [l_1, l_m]$ covers a fraction $\beta$ of the central points of $Z_{i,n-1}$ w.r.t. the calculated $L^p$ depth, i.e., it covers $R^\beta(Z_{i,n-1})$ for a given pre-set threshold $\beta \in (0,1)$. For both $X_t$ and $X_{t-1}$ we perform a simple binning procedure using the following bins: $(-\infty, l_1)$, $(l_1, l_2)$, ..., $(l_m, \infty)$. For **robust binning** we omit the "extreme" classes and use only the midpoints and bin frequencies for the classes $(l_1, l_2)$, $(l_2, l_3)$, ..., $(l_{m-1}, l_m)$.

Figs. 13 – 14 present the idea of simple $L^2$ binning in the case of data generated from a mixture of two two-dimensional normal distributions. The midpoints are represented by triangles.

Although, Hyndman and Yao in [11] and [12] considered a situation in which data were available in the form of a strictly stationary stochastic process $\{(X_i, Y_i)\}$, where $Y_i$ and $X_i$ are scalars, their estimators perform very well in the case of a *local DSA* as well. For the DSA, $X_i$ typically denotes a $k$ lagged value of $Y_i$. Let $g(y|x)$ be the conditional density of $Y_i$ given $X_i = x$. We are interested in robust estimation of $g(y|x)$ from the data $\{(X_l, Y_l), 1 \leq l \leq n\}$.

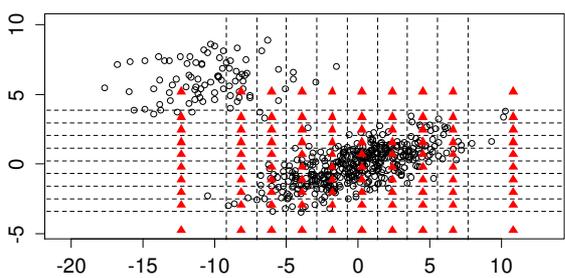
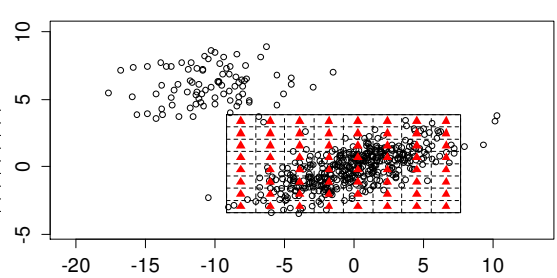

**Fig. 13:** The first step in $L^p$ depth binning.
*Source*: DepthProc R package.

**Fig. 14:** The second step in $L^p$ depth binning.
*Source*: DepthProc R package.

In our opinion, the best solution is a combination of robust binning and nonparametric estimation of the conditional density using the second proposal of Hyndman and Yao from [11], which they call a **constrained local polynomial estimator**.

Let

$$R(\boldsymbol{\theta}, x, y) = \frac{1}{B^{xy}} \sum_{i=1}^{m-1} \left\{ K_{hy}^{y}(Y_i' - y) - \sum_{j=0}^{r} \theta_j (X_i' - x)^j \right\}^2 \cdot K_{hx}^{x}(X_i' - x) b_i^{x'}, \qquad (20)$$

where $B^{xy} = \sum_{j=1}^{m-1} b_j^{y}$, $b_j^{y}$ are the "marginal" bin frequencies for $Y$, $Y_i'$, $X_j'$, denote the midpoints for the binned data $i, j = 1, \ldots, (m-1)$, and $b_i^{x'}$ denotes the frequency of $y$ under the condition that its $x$-component belongs to the same bin as $x'$. Then

$$\hat{g}(y \mid x) = \hat{\theta}_0, \qquad (21)$$

is a **local r-th order polynomial** estimator, where $\hat{\boldsymbol{\theta}}_{xy} = \left( \hat{\theta}_0, \hat{\theta}_1, \ldots, \hat{\theta}_r \right)^T$ is that value of $\boldsymbol{\theta}$ which minimizes $R(\boldsymbol{\theta}, x, y)$. Note that binning preserves properties of the **constrained local polynomial estimator.**

The estimator (21) uses two smoothing parameters: $hx$ controls the smoothness of conditional densities in the $x$ direction and $hy$ controls the smoothness of each conditional density in the $y$ direction. Effective (but computationally intensive) methods of the bandwidth selection were described in [11] and [12]. However, in the context of online analysis – we propose using a certain robust **rule of thumb** (see [26]). While estimator (21) has some nice properties, such as a smaller bias than the "classical" Nadaraya – Watson kernel density estimator when $r > 0$, it is not constrained to be non-negative and does not integrate to 1, except in the special case $r = 0$. For obtaining the nonnegativity Hyndman and Yao proposed setting

$$\hat{\theta}_0^* = l(\theta_0), \qquad (22)$$

where $l(u) = \exp(u)$.

The improved in this way estimator (22) considered jointly with the robust weighted $L^p$ binning is in our opinion the best for the purposes of DSA. In our opinion, it is enough to use polynomials of degree $r = 1, 2$.

**PROPOSAL 1**: Assume we analyze a stream $\{X_t\}$ using a moving window of fixed length $n$, i.e., $W_{i,n}$ and the derivative window $Z_{i,n-1}$. In the first step, we calculate the weighted sample $L^2$ depth for $W_{i,n}$. Next, we choose an equally spaced grid of points $l_1, \ldots, l_m$ in such way that $[l_1, l_m] \times [l_1, l_m]$ covers a fraction $\beta$ of the central points of $Z_{i,n-1}$ w.r.t. the calculated $L^2$ depth, i.e., it covers $R^\beta(Z_{i,n-1})$ for a certain pre-set threshold $\beta \in (0,1)$. For both $X_t$ and $X_{t-1}$ we perform a simple binning procedure using the following bins: $(-\infty, l_1)$, $(l_1, l_2)$,..., $(l_m, \infty)$. In the next step, we omit the "extreme" classes and to estimate the predictive distribution density function by means of (22) we use only the midpoints and binned frequencies for the classes $(l_1, l_2)$, $(l_2, l_3)$,..., $(l_{m-1}, l_m)$. For monitoring the PD of the stream, we use a minimum distance statistic of the form (13) using the *Hellinger* distance. We use bootstrap critical values based on the reference samples for making an intervention into the stream (or theoretical values when the reference densities are known).

The parameter *m* determines the degree of a "*sparsity*" of the binning and mainly relates to the window length and the computational complexity. We propose to take m=50–100 for windows of length 1000–10000 observations.

Note that $L^2$ depth is locally sensitive to outliers but, gives very robust estimators of centrality. We obtain a robust "support" for the binning, which rejects outliers, but stays sensitive to regime changes. This proposal protects us against outliers, but using the nearest neighbors bandwidth selection rule (e.g., offered by the *locfit* R package – see [26]), we can control the influence of inliers too. Note that it is possible to propose a **local robust binning** using an idea of local depth introduced in [21] and implemented in DepthProc [14] – this approach protects us against inliers.

**PROPOSAL 2**: Assume we analyze a multivariate stream $\{\mathbf{X}_t\}$ using a moving window of fixed length $n$, i.e., $\mathbf{W}_{i,n}$. To monitor the unconditional distribution of the stream, we calculate *moving multivariate Wilcoxon statistics* of the form (19) using $L^2$ depth and w.r.t. a fixed set of reference densities or samples. We use bootstrap critical values obtained w.r.t. reference samples do decide whether to make an intervention into the stream (or theoretical critical values in the case of known theoretical densities).

## VI. PROPERTIES OF THE PROPOSALS

It is worth noting several conceptual difficulties concerning understanding the robustness of a nonparametric estimator of a probability distribution. For example, if data are generated by a mixture of distributions, then a kernel density estimator tends to describe all the parts of the mixture, which could be treated as an advantage or disadvantage depending on one's point of view. In the DSA, using a "*majority voting*" rule, we focus our attention on the pattern represented by a majority of observations in the sample. This majority, however, can be defined by means of some global (protection against outliers) or local (protection against inliers) centrality measure (see [21])

To assess a breakdown of density estimator, we can take its unacceptable bias or variability at a fixed point, or use a given global measure, such as integrated mean squared error.

From a practical point of view, it is useful to evaluate the robustness of a density estimator in terms of the decision, for which it provides a basis. Our procedure breaks down, if it leads to only one decision, despite a continuum of possible samples and the possibility of multiple, regimes of the data stream (see [5]).

The quality of monitoring proposal 1 crucially depends on the quality of the density estimator used within the proposal. In order to assess performance of this proposal, we generated 500 samples of 1 000 000 obs. from several models of data streams having a strong practical justification. We estimated the CD based on windows of a fixed length of 500-50 000 obs. and considered samples without outliers and with up to 50% of additive outliers (AO) or inliers (IO) (see [19] for the definitions). We considered several CHARME schemes including one consist of two AR(1)-GARCH(1,1) sub-models

$$X_t = 5 + 0.1 X_{t-1} + \varepsilon_t, \varepsilon_t = \sigma_t Z_t, \sigma^2 = 1 + 0.1 \sigma_{t-1}^2 + 0.75 X_{t-1}^2, \qquad (23)$$

$$Y_t = 10 + 0.1 Y_{t-1} + \varepsilon_t, \varepsilon_t = \sigma_t Z_t, \sigma^2 = 1 + 0.1 \sigma_{t-1}^2 + 0.75 Y_{t-1}^2, \qquad (24)$$

where the innovations $\varepsilon_t$ come from a skewed Student distribution with 4 degrees of freedom, a skewed normal distribution, or skewed GED distribution (the default settings for the conditional distributions within the fGarch R package).

Our simulations also involved a CHARME scheme consisting of two SETAR models defined by

$$X_{t+1} = \begin{cases} 1 + 0.9 X_t + \varepsilon_{t+1} & X_{t-1} \leq 3 \\ 5 - 0.9 X_t + \varepsilon_{t+1} & X_{t-1} > 3 \end{cases}, \quad (25) \qquad Y_{t+1} = \begin{cases} 1 + 0.9 Y_t + \varepsilon_{t+1} & Y_{t-1} \leq 3 \\ 10 - 0.9 Y_t + \varepsilon_{t+1} & Y_{t-1} > 3 \end{cases}, \quad (26)$$

where the errors $\varepsilon_t$ were i.i.d. from the Student distribution with 3 degree of freedom.

We estimated the densities of the CD $\{X_t\}$ (Y) under the condition $\{X_{t-k} = a_l\}$ (X) for an equally spaced grid of 500 points from the interval [Med(sim) - A x MAD(sim), Med(sim)+ A x MAD(sim)], where Med(sim) and MAD(sim) are robust estimators of location and dispersion for the simulated trajectory based on, 20 equally spaced points $a_l$ representing the local conditions.

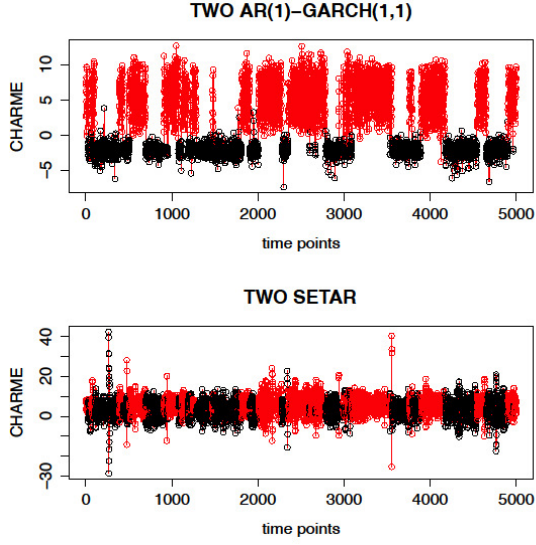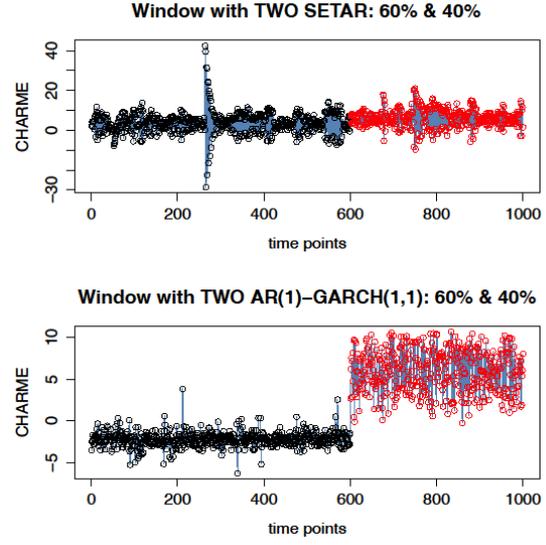

**Fig. 15:** Sample trajectories from CHARME models used in simulations.

**Fig. 16:** Windows consisted of points from two regimes of the CHARME models.

For each of the values of $X$ we condition on, we estimated the PD by means of proposal 1 and by means of the binned kernel density estimator (KERN) offered within KernSmooth package (a direct plug-in approach for bandwidth selection) package and by means of the default estimator offered by {hdrcde} – i.e., estimator (22) without binning (deg=1, link=log, method=1, bandwidth selection=AIC) (LOCPOL).

For each of the values of $X$ we condition on and for each consecutive time point, we calculated discrepancy measures between the estimated density and the known density (reference density) from the model used within the simulations at the time point:

$$R_1(\hat{g}, W_{i,n}) = \sum_{i=n_W+1}^{n_T} \underset{l}{MED}\{d_H(\hat{g}^i_{W_{i,N}}(y|X=a_l), f^i(y|X=a_l))\}, \tag{27}$$

$$R_2(\hat{g}, W_{i,n}) = \sum_{i=n_W+1}^{n_T} d_H(\hat{g}^i_{W_{i,n}}(y|X=x), f^i(y)), \tag{28}$$

where $n_W$ – denotes the window length, $n_T$ the number of considered time points, $l$ – the index of the value of $X$ which is being conditioned on, $d_H$ – the sum of absolute deviations between densities at the evaluation points $d_H = \sum_l |\hat{g}_l - f_l|$, $\hat{g}^i_{W_{i,n}}$ estimated density, $f^i$ true density, $MED$ – the median.

We considered data generating schemes differing w.r.t. the transition matrix of the CHARME.

Tab. 1 presents averaged sums of absolute deviations $d_H$ between the true CD distribution for various models for generating the data with a set proportion of outliers (varying from 0 to 10%) and selected estimators of the conditional density: the KERN, the LOCPOL and proposal 1 (PROP 1) for windows consisting of 10 000 obs. generated from the CHARME model consisting of two AR(1)-GARCH(1,1) sub-models defined by (23) and (24). We only conditioned on one value of $X$. The parameters used for proposal 1 were fixed as $m = 200$, $\beta = 0.05$. We considered windows consisting of 10-40% obs. from the first sub-model and the rest from the second sub-model. The windows consisted of up to 45% of outliers and inliers generated from a mixture of 7 normal distributions, where six of them had supports concentrated in the central part of the unconditional CHARME distribution and one of them had a ten times larger variance than the variance of the simulated data. Tab. 2 presents analogous results to Tab. 1 in the case of the CHARME model consisting of two SETAR sub-models defined by (25) and (26)

Although we observed a relatively high dispersion of the simulated discrepancy measures – the general tendency is in favor of our proposal. The high quality of our proposals starts to be evident with outlier fraction exceeding 10%. The behavior of the proposals based on the whole of the simulated trajectories using criteria (27) and (28) was also very good. However, since the quality of the density estimator is of prime importance for monitoring proposal 1, we studied the small sample behavior of statistic (11) for the *Hellinger* and *Kolmogorov* distances (see [25] for properties of this statistic) and the CD density estimator with robust binning. The simulation studies were at least very promising for our proposal in comparison to other parametric, as well nonparametric, density estimators. The estimated computational time for a window consisting of computation for window consisted of 1 000 obs. was 1.46sec for binned kernel KERN, 3.8sec for LOC and 0.84sec for PROP 1. For a window consisting of 10 000 obs., we observed a computational time 1.34sec for binned KERN, 3min 34sec for LOC, and 1min 47 sec for PROP 1. We used the KernSmooth R package for kernel estimation, the hdrcde package for the constrained local estimator, and implemented proposal 2 (binningDepth2D within the DepthProc) with 100x100 binning.

Proposal 2 is mainly adapted to monitoring multivariate data streams. To check its properties, we studied several simulation schemes involving CHARME models consisting of vector autoregressive models (VARs) and multivariate GARCH models. We studied the behavior of statistic (21), as well as the usefulness of the *moving DD-plot*. The results of the small sample studies were very promising. Fig. 17 presents 5-min quotations for 5 stocks belonging to the Dow Jones Industrial Index in the period from 2008-03 to 2013-03. Figure 18 presents the first differences for these quotations. Fig. 19 presents an application to the time series from Fig. 17, the moving multivariate Wilcoxon statistic (19) calculated from 100-elements window with the reference sample taken to be the first 100 observations.

**Tab 1**. Performance of the kernel PD estimator (KERN), constrained local polynomial estimator (LOCPOL), the estimator from proposal 1 (PROP 1) for windows consisting of 1 000 obs. generated from the sub-models defined by (27) and (28). The table consists of the mean values of $d_H$ from 100 repetitions.

| 2 x SETAR | KERN | LOCPOL | PROP 1 |
|---|---|---|---|
| 10%sub1-90%sub2 | 9.41 | 9.29 | 7.29 |
| 20%sub1-80%sub2 | 7.67 | 8.32 | 6.60 |
| 30%sub1-70%sub2 | 10.11 | 10.63 | 8.61 |
| 40%sub1-60%sub2 | 11.27 | 10.89 | 7.75 |
| 10%-90%+5%AO | 5.44 | 5.35 | 4.29 |
| 20%-80%+5%AO | 10.99 | 10.60 | 8.56 |
| 30%-70%+5%AO | 12.52 | 12.1 | 10.88 |
| 40%-60%+5%AO | 6.33 | 6.22 | 5.00 |
| 10%-90%+10%AO | 4.56 | 4.47 | 3.73 |
| 20%-80%+10%AO | 9.78 | 10.31 | 7.75 |
| 30%-70%+10%AO | 10.84 | 10.68 | 8.65 |
| 40%-60%+10%AO | 8.38 | 8.27 | 6.34 |

*Source: Our own calculations, DepthProc package.*

**Tab 2**. Performance of the kernel PD estimator (KERN), constrained local polynomial estimator (LOCPOL), the estimator from proposal 1 (PROP 1) for windows consisting of 1 000 obs. generated from the sub-models defined by (25) and (26). The table consists of the mean values of $d_H$ from 100 repetitions.

| 2 x AR-GARCH | KERN | LOCPOL | PROP 1 |
|---|---|---|---|
| 10%sub1-90%sub2 | 3.28 | 3.69 | 3.48 |
| 20%sub1-80%sub2 | 7.99 | 7.97 | 6.11 |
| 30%sub1-70%sub2 | 14.97 | 15.59 | 13.14 |
| 40%sub1-60%sub2 | 24.37 | 25.09 | 23.39 |
| 10%-90%+5%AO | 4.13 | 6.58 | 5.26 |
| 20%-80%+5%AO | 5.69 | 5.68 | 5.25 |
| 30%-70%+5%AO | 7.16 | 7.17 | 6.8 |
| 40%-60%+5%AO | 13.39 | 13.07 | 12.12 |
| 10%-90%+10%AO | 3.94 | 4.97 | 4.55 |
| 20%-80%+10%AO | 4.01 | 4.00 | 3.79 |
| 30%-70%+10%AO | 6.56 | 6.57 | 6.32 |
| 40%-60%+10%AO | 8.18 | 8.19 | 8.01 |

*Source: Our own calculations, DepthProc package.*

Fig. 20 presents an analogous situation, but the Wilcoxon statistic is applied to the time series from Fig. 18. Red lines on Fig. 19 – 20 present certain threshold fixed by analyst for intervention purposes. It is easy to notice that proposal 2 helps us in detecting a change trend or a scale change in within the multidimensional stream. Additionally, intensive simulation studies confirm the clear merits of moving DD-plot based statistics in the context of DSA monitoring.

Note, that both proposals are robust to a small fraction of outliers and they are sensitive to regime changes at the same time. A full description of the simulation results may be found in [17].

## VII. CONCLUSSIONS

We presented two robust procedures adapted to the analysis of streaming data. These procedures behave very well in cases of data which have a moderate fraction of outliers in comparison to procedures based on classical approaches to statistical inference. The R implementation of the proposals is completely freely available on CRAN servers in the DepthProc R package. These procedures are still being developed in the context of the possibilities of distributed and and/or recursive inference online. For robust procedures, this is problem of prime importance, unsolved as far, particularly in the multivariate case. Certain proposals in this matter can be found in [13] and [17].

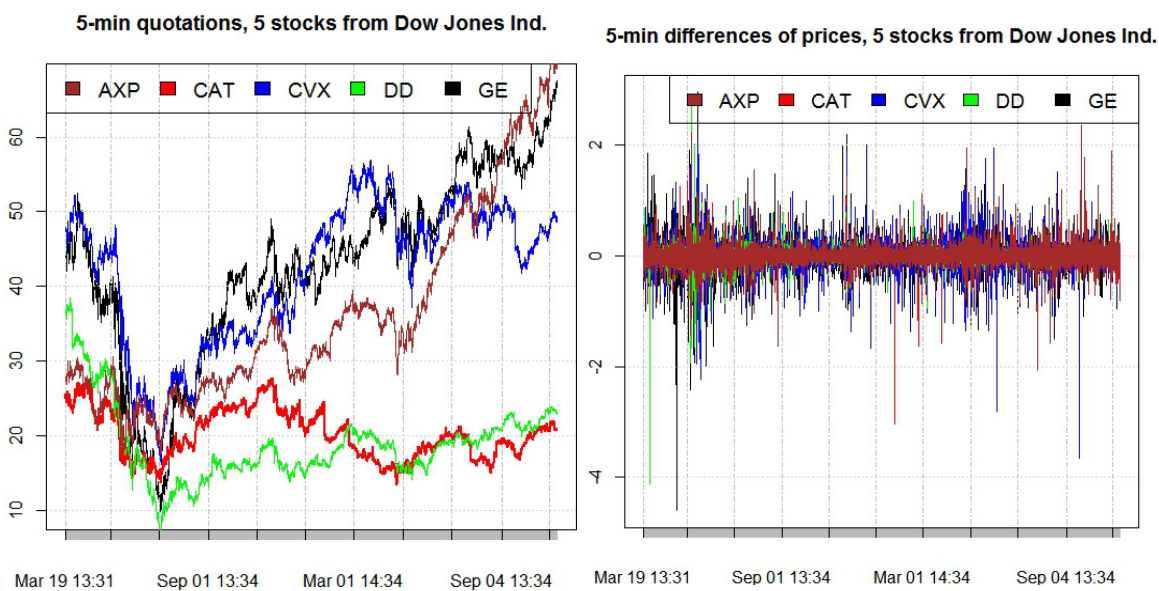

**Fig. 17** 5-min quotations for stocks from Dow Jones Ind. *Source: Our own calculations.*

**Fig. 18** First differences 5-min quotations for stocks from DJ Ind. *Source: Our own calculations*

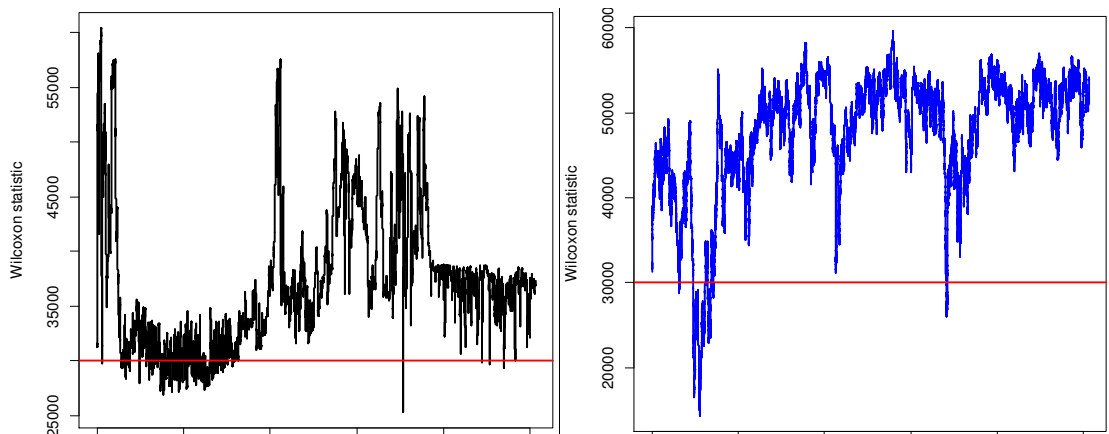

**Fig. 19** Moving Wilcoxon statistics from 250-obs window, 5-min quotations for stocks from Dow Jones Ind. *Source: DepthProc R package.*

**Fig. 20** Moving Wilcoxon statistics from 250-obs window, first differences of 5-min quotations for stocks from Dow Jones Ind. *Source: DepthProc.*

ACKNOWLEDGMENT

The author thanks the Polish National Science Center for financial support from grant UMO-2011/03/B/HS4/01138.